\documentstyle[twoside,epsfig]{article}

\catcode`\@=11
\long\def\@makefntext#1{
\protect\noindent \hbox to 3.2pt {\hskip-.9pt  
$^{{\eightrm\@thefnmark}}$\hfil}#1\hfill}               

\def\thefootnote{\fnsymbol{footnote}}
\def\@makefnmark{\hbox to 0pt{$^{\@thefnmark}$\hss}}    
        
\def\ps@myheadings{\let\@mkboth\@gobbletwo
\def\@oddhead{\hbox{}
\rightmark\hfil\eightrm\thepage}   
\def\@oddfoot{}\def\@evenhead{\eightrm\thepage\hfil
\leftmark\hbox{}}\def\@evenfoot{}
\def\sectionmark##1{}\def\subsectionmark##1{}}

\oddsidemargin=\evensidemargin
\renewcommand{\thefootnote}{\fnsymbol{footnote}}

\newcounter{sectionc}\newcounter{subsectionc}\newcounter{subsubsectionc}
\renewcommand{\section}[1] {\vspace{12pt}\addtocounter{sectionc}{1} 
\setcounter{subsectionc}{0}\setcounter{subsubsectionc}{0}\noindent 
        {\tenbf\thesectionc. #1}\par\vspace{5pt}}
\renewcommand{\subsection}[1] {\vspace{12pt}\addtocounter{subsectionc}{1} 
        \setcounter{subsubsectionc}{0}\noindent 
        {\bf\thesectionc.\thesubsectionc. {\kern1pt \bfit #1}}\par\vspace{5pt}}
\renewcommand{\subsubsection}[1] {\vspace{12pt}\addtocounter{subsubsectionc}{1}
        \noindent{\tenrm\thesectionc.\thesubsectionc.\thesubsubsectionc.
        {\kern1pt \tenit #1}}\par\vspace{5pt}}
\newcommand{\nonumsection}[1] {\vspace{12pt}\noindent{\tenbf #1}
        \par\vspace{5pt}}

\newcounter{appendixc}
\newcounter{subappendixc}[appendixc]
\newcounter{subsubappendixc}[subappendixc]
\renewcommand{\thesubappendixc}{\Alph{appendixc}.\arabic{subappendixc}}
\renewcommand{\thesubsubappendixc}
        {\Alph{appendixc}.\arabic{subappendixc}.\arabic{subsubappendixc}}

\renewcommand{\appendix}[1] {\vspace{12pt}
        \refstepcounter{appendixc}
        \setcounter{figure}{0}
        \setcounter{table}{0}
        \setcounter{lemma}{0}
        \setcounter{theorem}{0}
        \setcounter{corollary}{0}
        \setcounter{definition}{0}
        \setcounter{equation}{0}
        \renewcommand{\thefigure}{\Alph{appendixc}.\arabic{figure}}
        \renewcommand{\thetable}{\Alph{appendixc}.\arabic{table}}
        \renewcommand{\theappendixc}{\Alph{appendixc}}
        \renewcommand{\thelemma}{\Alph{appendixc}.\arabic{lemma}}
        \renewcommand{\thetheorem}{\Alph{appendixc}.\arabic{theorem}}
        \renewcommand{\thedefinition}{\Alph{appendixc}.\arabic{definition}}
        \renewcommand{\thecorollary}{\Alph{appendixc}.\arabic{corollary}}
        \noindent{\tenbf Appendix \theappendixc #1}\par\vspace{5pt}}
\newcommand{\subappendix}[1] {\vspace{12pt}
        \refstepcounter{subappendixc}
        \noindent{\bf Appendix \thesubappendixc. {\kern1pt \bfit #1}}
        \par\vspace{5pt}}
\newcommand{\subsubappendix}[1] {\vspace{12pt}
        \refstepcounter{subsubappendixc}
        \noindent{\rm Appendix \thesubsubappendixc. {\kern1pt \tenit #1}}
        \par\vspace{5pt}}

\newcommand{\textlineskip}{\baselineskip=13pt}
\newcommand{\smalllineskip}{\baselineskip=10pt}

\def\eightcirc{
\begin{picture}(0,0)
\put(4.4,1.8){\circle{6.5}}
\end{picture}}
\def\eightcopyright{\eightcirc\kern2.7pt\hbox{\eightrm c}} 

\newcommand{\copyrightheading}[1]
        {\vspace*{-2.5cm}\smalllineskip{\flushleft
        {\footnotesize International Journal of Modern Physics C #1}\\
        {\footnotesize $\eightcopyright$\, World Scientific Publishing
         Company}\\
         }}

\def\abstracts#1#2#3{{
        \centering{\begin{minipage}{4.5in}\footnotesize\baselineskip=10pt
        \parindent=0pt #1\par 
        \parindent=15pt #2\par
        \parindent=15pt #3
        \end{minipage}}\par}} 

\def\keywords#1{{
        \centering{\begin{minipage}{4.5in}\footnotesize\baselineskip=10pt
        {\footnotesize\it Keywords}\/: #1
         \end{minipage}}\par}}

\newcommand{\bibit}{\nineit}
\newcommand{\bibbf}{\ninebf}
\renewenvironment{thebibliography}[1]
        {\frenchspacing
         \ninerm\baselineskip=11pt
         \begin{list}{\arabic{enumi}.}
        {\usecounter{enumi}\setlength{\parsep}{0pt}     
         \setlength{\leftmargin 12.7pt}{\rightmargin 0pt} 
         \setlength{\itemsep}{0pt} \settowidth
        {\labelwidth}{#1.}\sloppy}}{\end{list}}

\newcounter{itemlistc}
\newcounter{romanlistc}
\newcounter{alphlistc}
\newcounter{arabiclistc}

\newcommand{\fcaption}[1]{
        \refstepcounter{figure}
        \setbox\@tempboxa = \hbox{\footnotesize Fig.~\thefigure. #1}
        \ifdim \wd\@tempboxa > 5in
           {\begin{center}
        \parbox{5in}{\footnotesize\smalllineskip Fig.~\thefigure. #1}
            \end{center}}
        \else
             {\begin{center}
             {\footnotesize Fig.~\thefigure. #1}
              \end{center}}
        \fi}

\newcommand{\tcaption}[1]{
        \refstepcounter{table}
        \setbox\@tempboxa = \hbox{\footnotesize Table~\thetable. #1}
        \ifdim \wd\@tempboxa > 5in
           {\begin{center}
        \parbox{5in}{\footnotesize\smalllineskip Table~\thetable. #1}
            \end{center}}
        \else
             {\begin{center}
             {\footnotesize Table~\thetable. #1}
              \end{center}}
        \fi}

\def\@citex[#1]#2{\if@filesw\immediate\write\@auxout
        {\string\citation{#2}}\fi
\def\@citea{}\@cite{\@for\@citeb:=#2\do
        {\@citea\def\@citea{,}\@ifundefined
        {b@\@citeb}{{\bf ?}\@warning
        {Citation `\@citeb' on page \thepage \space undefined}}
        {\csname b@\@citeb\endcsname}}}{#1}}

\newif\if@cghi
\def\cite{\@cghitrue\@ifnextchar [{\@tempswatrue
        \@citex}{\@tempswafalse\@citex[]}}
\def\citelow{\@cghifalse\@ifnextchar [{\@tempswatrue
        \@citex}{\@tempswafalse\@citex[]}}
\def\@cite#1#2{{$\null^{#1}$\if@tempswa\typeout
        {IJCGA warning: optional citation argument 
        ignored: `#2'} \fi}}

\def\pmb#1{\setbox0=\hbox{#1}
        \kern-.025em\copy0\kern-\wd0
        \kern.05em\copy0\kern-\wd0
        \kern-.025em\raise.0433em\box0}

\def\fnt#1#2{\footnotetext{\kern-.3em
        {$^{\mbox{\scriptsize #1}}$}{#2}}}

\def\ps@myheadings{%
    \let\@oddfoot\@empty\let\@evenfoot\@empty
    \def\@evenhead{\slshape\leftmark\hfil}
    \def\@oddhead{\hfil{\slshape\rightmark}}
    \let\@mkboth\@gobbletwo
    \let\sectionmark\@gobble
    \let\subsectionmark\@gobble
    }
\font\tenrm=cmr10
\font\tenit=cmti10 
\font\tenbf=cmbx10
\font\bfit=cmbxti10 at 10pt
\font\ninerm=cmr9
\font\nineit=cmti9
\font\ninebf=cmbx9
\font\eightrm=cmr8

\def\qed{\hbox{${\vcenter{\vbox{                    
   \hrule height 0.4pt\hbox{\vrule width 0.4pt height 6pt
   \kern5pt\vrule width 0.4pt}\hrule height 0.4pt}}}$}}

\renewcommand{\thefootnote}{\fnsymbol{footnote}}    

\def\bsc{{\sc a\kern-6.4pt\sc a\kern-6.4pt\sc a}}       
\def\bflatex{\bf L\kern-.30em\raise.3ex\hbox{\bsc}\kern-.14em 
T\kern-.1667em\lower.7ex\hbox{E}\kern-.125em X} 

\begin{document}
\setlength{\textheight}{7.7truein}  

\thispagestyle{empty}

\markboth{\protect{\footnotesize\it Multidimensional Consensus Model
on a Barab\'asi-Albert Network}}{\protect{\footnotesize\it Multidimensional Consensus Model
on a Barab\'asi-Albert Network}}

\normalsize\textlineskip

\setcounter{page}{1}

\copyrightheading{}                     

\vspace*{0.88truein}

\centerline{\bf MULTIDIMENSIONAL CONSENSUS MODEL}
\vspace*{0.035truein}
\centerline{\bf ON A BARAB\'ASI-ALBERT NETWORK}
\vspace*{0.37truein}
\centerline{\footnotesize DIRK JACOBMEIER}
\baselineskip=12pt
\centerline{\footnotesize\it Institute for Theoretical Physics, Cologne 
University}
\baselineskip=10pt
\centerline{\footnotesize\it 50923 K\"oln, Germany}
\centerline{\footnotesize\it E-mail: dj@thp.uni-koeln.de}

\vspace*{0.25truein}
\abstracts{
A Consensus Model according to Deffuant on a directed
Barab\'asi-Albert network was simulated.
Agents have opinions on different subjects.
A multi-component subject vector was used.
The opinions are discrete. The analysis regards distribution
and clusters of agents which are on agreement in the opinions 
of the subjects. 
Remarkable results are on the one hand, that there mostly 
exists no absolute consens. It determines depending on the ratio of
number of agents to the number of subjects,
whether the communication ends in a consens or a pluralism.
Mostly a second robust cluster remains, in its size depending
on the number of subjects.
Two agents agree either in (nearly) all or (nearly) no subject.\\
The operative parameter of the consens-formating-process is the
tolerance in change of views of the group-members.}{}{}

\vspace*{5pt}

\setcounter{footnote}{0}
\renewcommand{\thefootnote}{\alph{footnote}}

\vspace*{1pt}\textlineskip    

\section{Introduction} 
\vspace*{-0.5pt}
\noindent
"Winwood Reade is good upon the subject," said Holmes. "He remarks that, 
while the individual man is an insoluble puzzle, in the aggregate he 
becomes a mathematical certainty. You can, for example, never foretell 
what any one man will do, but you can say with precision what an average 
number will be up to. Individuals vary, but percentages remain constant. 
So says the statistician. ..."\\
{The Sign of the Four; Arthur Conan Doyle, 1890}\\
\\
The enigma of man shouldn't be solved here. It is our aim to build a model, 
which imitates the behaviour of a group of humans.
Therefore from the behaviour of people rules have been deduced
which were placed into models \cite{1,2,3}.\\
Consensus models recreate the opinion forming process of a group
of agents. 
One starts from a random distribution of opinions.  
After a simulation of the communication of agents with each other,
the resulting distribution of opinions is regarded. 
There have been developed and simulated some consensus models in
which the choice of connections of the agents among each other
is multifaceted.
Models differ in the used topologies, in their way of
communication, the relationship of agents to 
a subject, the dimensionality of geometry, subject, and opinion, etc.
\cite{4,5,6,7,8,9}.\\ 
The present model continues this tradition, and offers a 
consensus model according to Deffuant \cite{7} on a directed
Barab\'asi-Albert network \cite{10} with discrete opinions and several 
subjects (\,= questions, themes, ... ).
Every agent $i$ ($i\,=\,1,2,...,N$) has on each subject $S_k$ 
($k\,=\,1,2,...,S$) an opinion $O_i^k$.
The discrete opinion spectrum comprises natural numbers from 1 to O.\\
Simulations of a consensus model \'a la Deffuant on a directed
Barab\'asi-Albert network with one subject and discrete opinions have been 
made by Stauffer et al.\cite{11}.

\section{Model}
\vspace*{-0.5pt}

\subsection{Network assembly}
\noindent
At the beginning one knot of $m$ agents, each connected with all others,
is built. 
Every newly added agent $i$ connects itself with
$m$ already existing agents in the network. The connection 
takes place stochastically. With it the probability of connecting
with a already existing agent is proportional to the total number of 
its connections (``The rich get richer'').\\ Besides
the connection is directed, i.e., the agents communicate 
along the $m$ connections, which they assemble themselves. The
connections, with whom they get in touch when new agents are added,
can not be chosen by themselves.

\subsection{Communication}
\noindent
The communication takes place along the connections. The agents become
the active communicator $i$ in the order they have been bound into 
the network.
The partner for communication $j$ will be chosen randomly from the 
$m$ with those to whom $i$ has connected itself.
Then the over-all distance $\delta$ to the partner of communication will
be calculated. This $\delta$ results from the absolute value of the distance
of all opinions to each other
\begin{equation}
\delta\,=\,\sum_{k=1}^S\,|O_i^k\,-\,O_j^k|\,, \label{math1}
\end{equation}
and is the indicator for the start of a communication:
\begin{itemize}
\item [-]If $\delta$ greater than a given $\Delta=\,O\,S\,\varepsilon$\,
a communication will be impossible and ends for this agent, where 
$\varepsilon$ with $0\,<\,\varepsilon\,<\,1$ is an input parameter.
Then it is the next agents' turn.
\item [-]If $\delta$ is lower or equal the given $\Delta$ then a 
communication will start.
\end{itemize}
The outcomes of a simulation with $\Delta=\,O\,S\,\varepsilon$
don't differ in the substance compared with the outcomes of a simulation
with $\Delta=\,(O-1)\,S\,\varepsilon$. The typical appearances of the
model remain preserved.

\subsubsection{Rules for Simulating the Communication:}
\noindent
Now agents $i$ and $j$ look randomly for a subject $S_k$ on which they 
will communicate. 
\begin{itemize}
\item If the difference of opinions $(O_i^k\,-\,O_j^k)$ of both partners of
communication on the subject $k$ results in zero, then they 
agreeing and the communication ends.
\item If the difference of opinions equals one, one communicant will
adopt randomly the opinion from the other.
\item If the difference of opinions is larger than one, both communicants
approach each other about $d$, with rounding the opinion.\\
With $d\,=\,\sqrt{1/10}\,\,\,(O_i^k\,-\,O_j^k)\,$, it will be 
$O_i^k\,:=\,O_i^k\,-\,d$ and $O_j^k\,:=\,O_j^k\,+\,d$.
\end{itemize}
After that it is the next agents turn.\\
The simulation ends, when during $n$ iterations over all agents no 
change of opinion in one of the communications takes place.  

\subsection{Parameter}
\noindent
The parameters of the model, which have been modified, are \,\,
$N$: Number of agents;\,\,
$S$: Number of subjects;\,\,
$O$: Number of opinions per subject;\,\,
$\varepsilon$: tolerance,\\ $\Delta=\,O\,S\,\varepsilon$, 
$0\,\le\varepsilon\le\,1$;\,\,
$n$: stop criterion; the simulation stops if during $n$ consecutive 
iterations over all agents no opinion was changed. 

\subsection{Methods of Evaluation}
\begin{figure}[ht!]
\vspace*{13pt}
\centerline{\epsfig{file=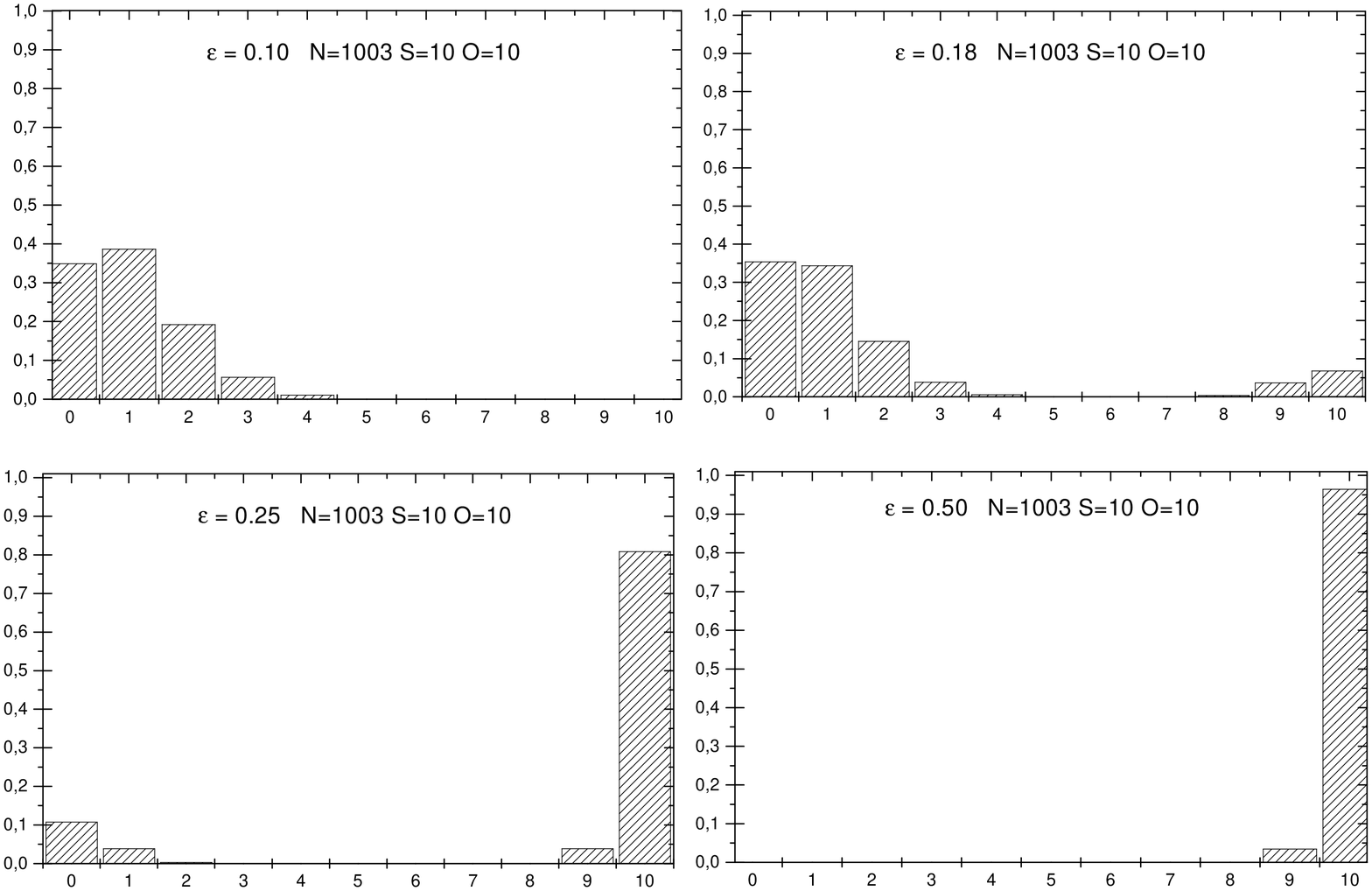, scale=0.45}}
\vspace*{13pt}
\fcaption{Distribution for $S\,=\,10$: Quota of pairs of 
agents agreeing on the plotted number of subjects.
With growing tolerance $\varepsilon$, the number of maximal
agreeing agents grows. The places of the distribution with a 
intermediate number of agreements remain vacant 
(resp. very sparsely populated).
The opinion of agents matching in all subjects lies in the centrist 
opinion of the opinion spectrum $O$ (here with $O$\,=\,10 on the 
opinion 5 or 6 each) in every subject.}
\label{Vert}
\end{figure}

\noindent
\begin{itemize}
\item [a)]{\bf Distribution}\\ The distribution specifies how many persons
in how many subjects share the same opinion. The quota will be plotted
as a fraction of the total number of all agents.\\ 
At the beginning the shape of distribution resembles a random distribution.
\item[b)]{\bf ``Center of Mass''}\\The ``Center of Mass'' of the 
distribution is the average number of 
subjects on which one agents agree with one another and results
from the weighted distribution (cp. \ref{Vert}: For the ``Center of Mass'' 
the shaded area of the figure is considered).\\
The "Center of Mass" is analogous to the order parameter of Klemm et al.
\cite{8}.
\item[c)]{\bf Cluster} \\A cluster contains all agents in the network,
who share the same opinion on any of the $S$ subjects.
\end{itemize}

\section{Simulation}
\vspace*{-0.5pt}

\subsection{Description}
\noindent 
The tolerance $\varepsilon$ was varied.
At very small $\varepsilon$ the distribution (Fig.\,\ref{Vert}) 
is on the left part a combination of random distributions.
The number of agents on the left side decreases with growing $\varepsilon$.
The number of agents on the right side (agreeing in (nearly) all subjects) 
increases to the same extent.
\begin{figure}[ht!]
\centerline{\psfig{file=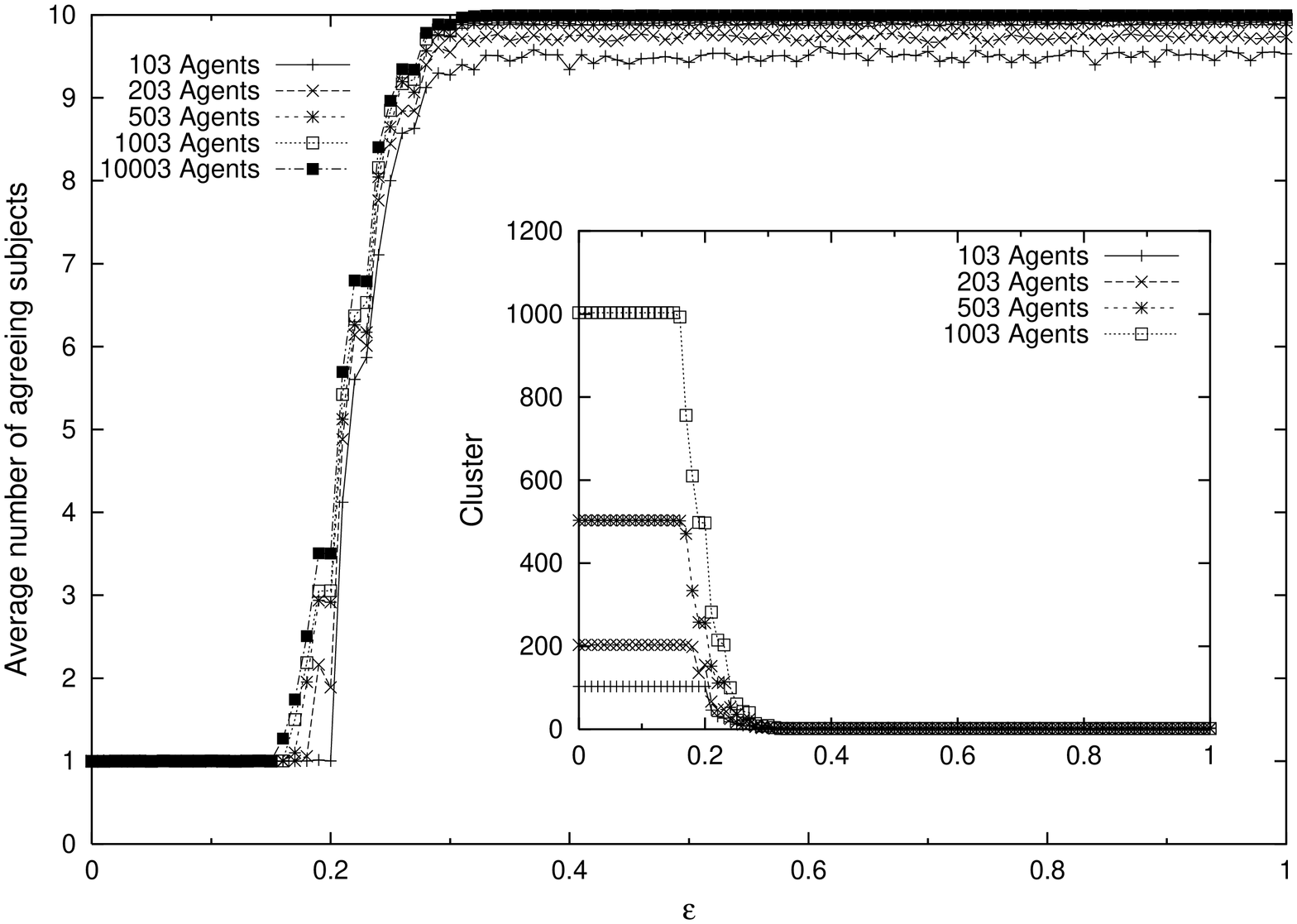, 
scale=0.52, angle=-90}}
\fcaption{The ``Center of Mass'' of the distribution vs.
$\varepsilon$ is plotted. Dependent on $\varepsilon$ the ``Center of Mass'' 
and number of clusters vary very fast (see inset). Within a small 
$\varepsilon$-interval from $\varepsilon_1$ to $\varepsilon_2$ the 
maximum of ``Center of Mass'' and the minimal
number of clusters is reached.  The corresponding $\varepsilon$-interval
varies with the total number of agents $N$.}
\label{Schwere}
\end{figure}
Two peculiarities are to take note of. On the one hand, no agents 
are found, which agree in an intermediate number of subjects.
The places of the distribution with an intermediate number of 
agreements remain vacant.
The extremal distribution of accordance and dissent are favoured. 
On the other hand, there is no ``absolute consensus'' 
(an ``absolute consensus'' is an accordance of all agents in all subjects).
There remain for $\varepsilon > \varepsilon_2$ 
($\varepsilon_2$ is the $\varepsilon$ above which the average number 
of clusters remains stable) several (typically 2) clusters.\\   
The other $\varepsilon$ below which the number of clusters
is constant (see inset Fig.\,\ref{Schwere}), is called $\varepsilon_1$.

\subsection{Analysis}
\noindent
$N$, $S$ and $O$ have been varied, as outlined.

\subsubsection{$N$}
\begin{figure}[ht!]
\centerline{\psfig{file=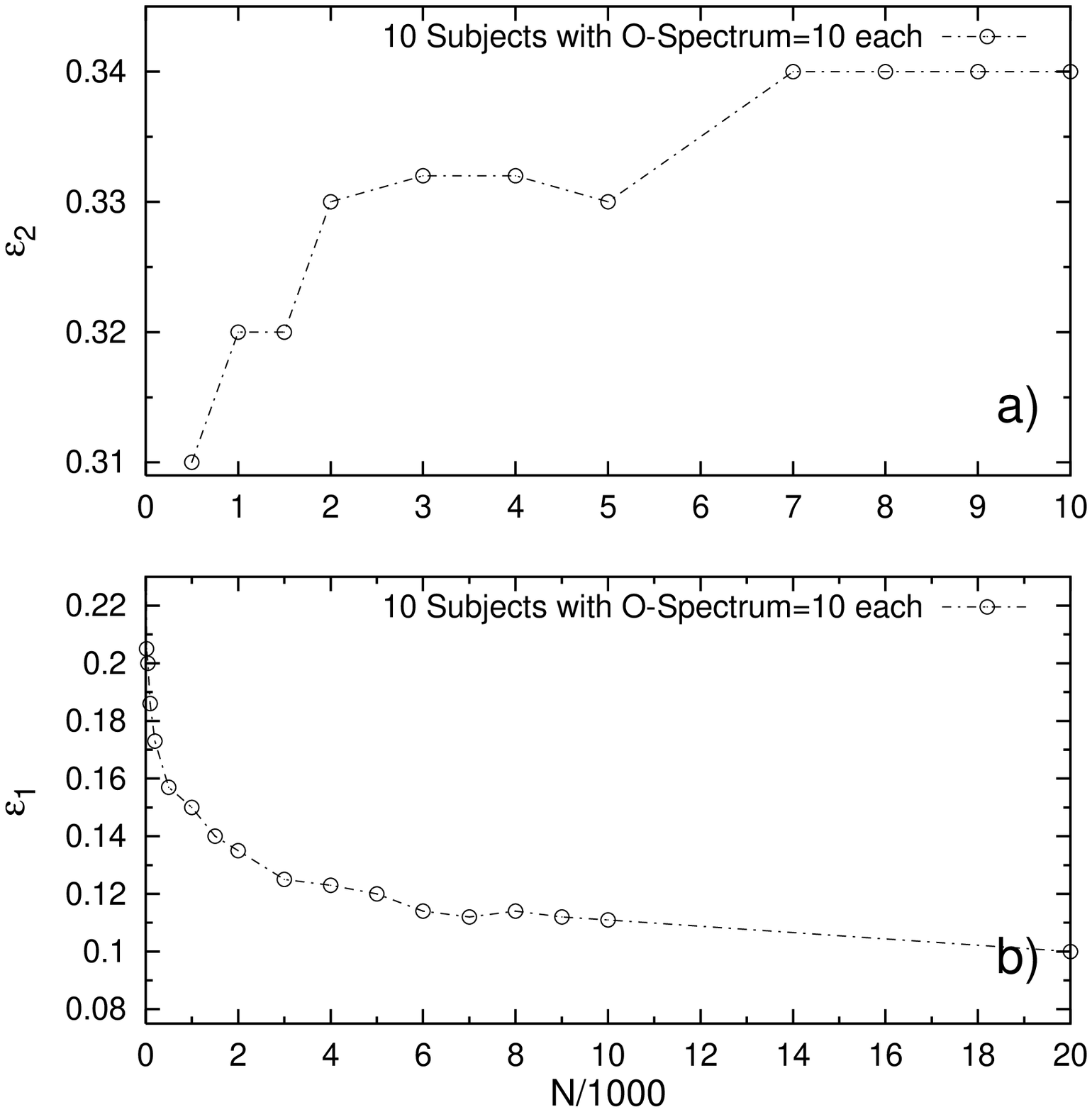, scale=0.7, angle=-90}}
\fcaption{a) shows the variation of $\varepsilon_2$ with the number
$N$ of agents.\\
\hspace*{0.98cm}b) shows the $N$-variation of $\varepsilon_1$.}
\label{Nvar}
\end{figure}

\noindent
With increasing $N$, $\varepsilon_1$ decreases and
the number of clusters as well as $\varepsilon_2$ increase, 
Figure \ref{Nvar}.\\  
In a social group of many people they sooner find each other
for a discussion, and thereby starting a process of agreement.   
As $\varepsilon_2$ is weakly depending on the number of agents,
its change with $N$ for small groups is higher (Fig.\,\ref{Nvar}).\\

\subsubsection{$S$}
\begin{figure}[ht!]
\centerline{\psfig{file=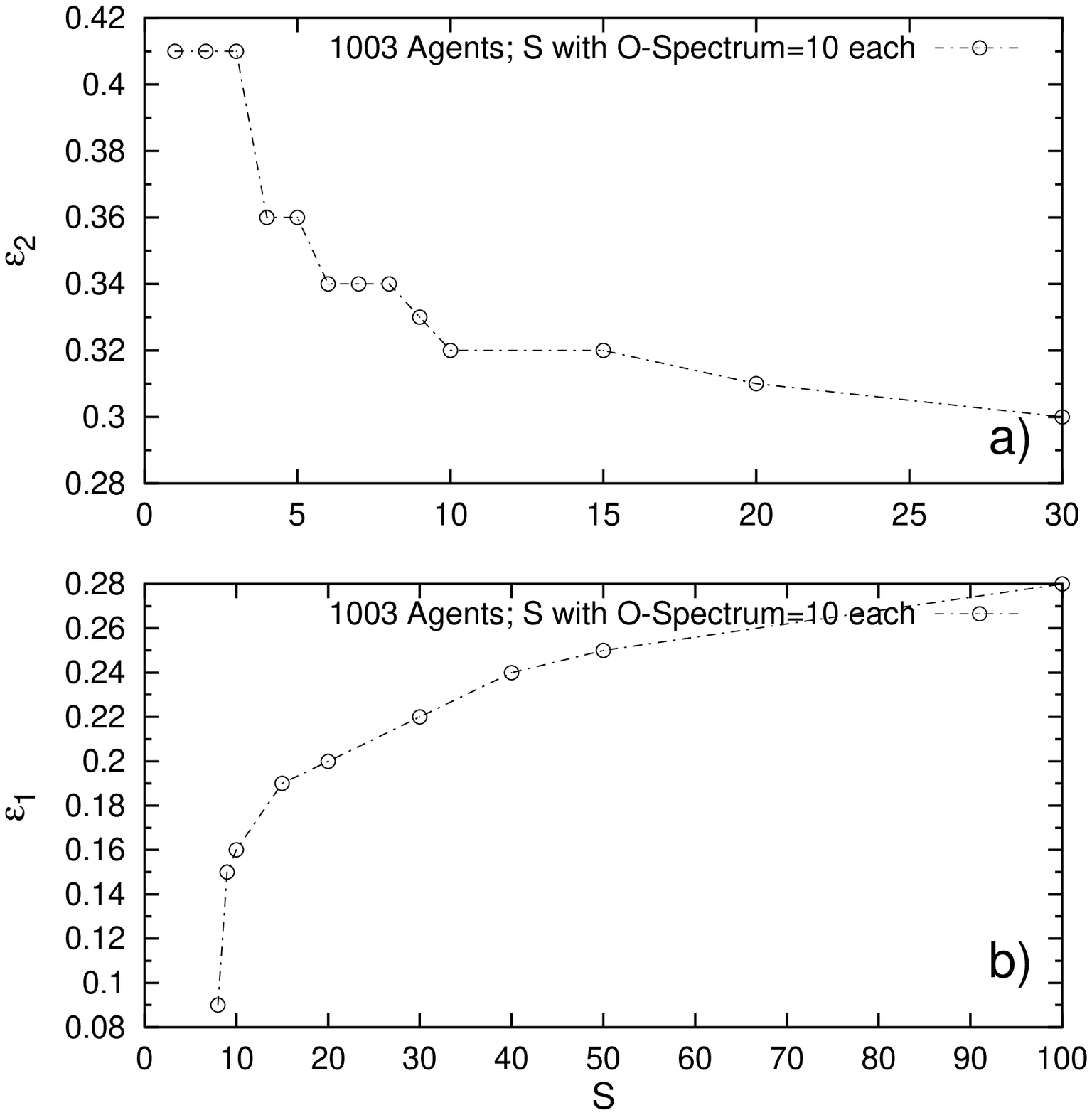, scale=0.7, angle=-90}}
\fcaption{a) $\varepsilon_2$ and its variation with $S$.\\
\hspace*{0.98cm}b) $\varepsilon_1$ and its variation with $S$.}
\label{Svar}
\end{figure}

\noindent
$\varepsilon_1$ and $\varepsilon_2$ (Fig.\,\ref{Svar}\,a, b\,) 
also depend on the number of the discussed subjects $S$.\\
A large $S$ needs a larger $\varepsilon_1$ to start discussion, 
compared with small $S$ (Fig.\,\ref{Svar}\,b{\small )}\,). 
I.e., if many subjects are available the communicants 
need larger $\varepsilon_1$ to start discussion, compared with small $S$.\\
The less subjects are available for selection the 
smaller is the $\varepsilon_1$ where a talk begins.
But then it needs a larger tolerance of the people speaking together,
before an agreement will be found. 

\subsubsection{$O$}
\begin{figure}[ht!]
\centerline{\psfig{file=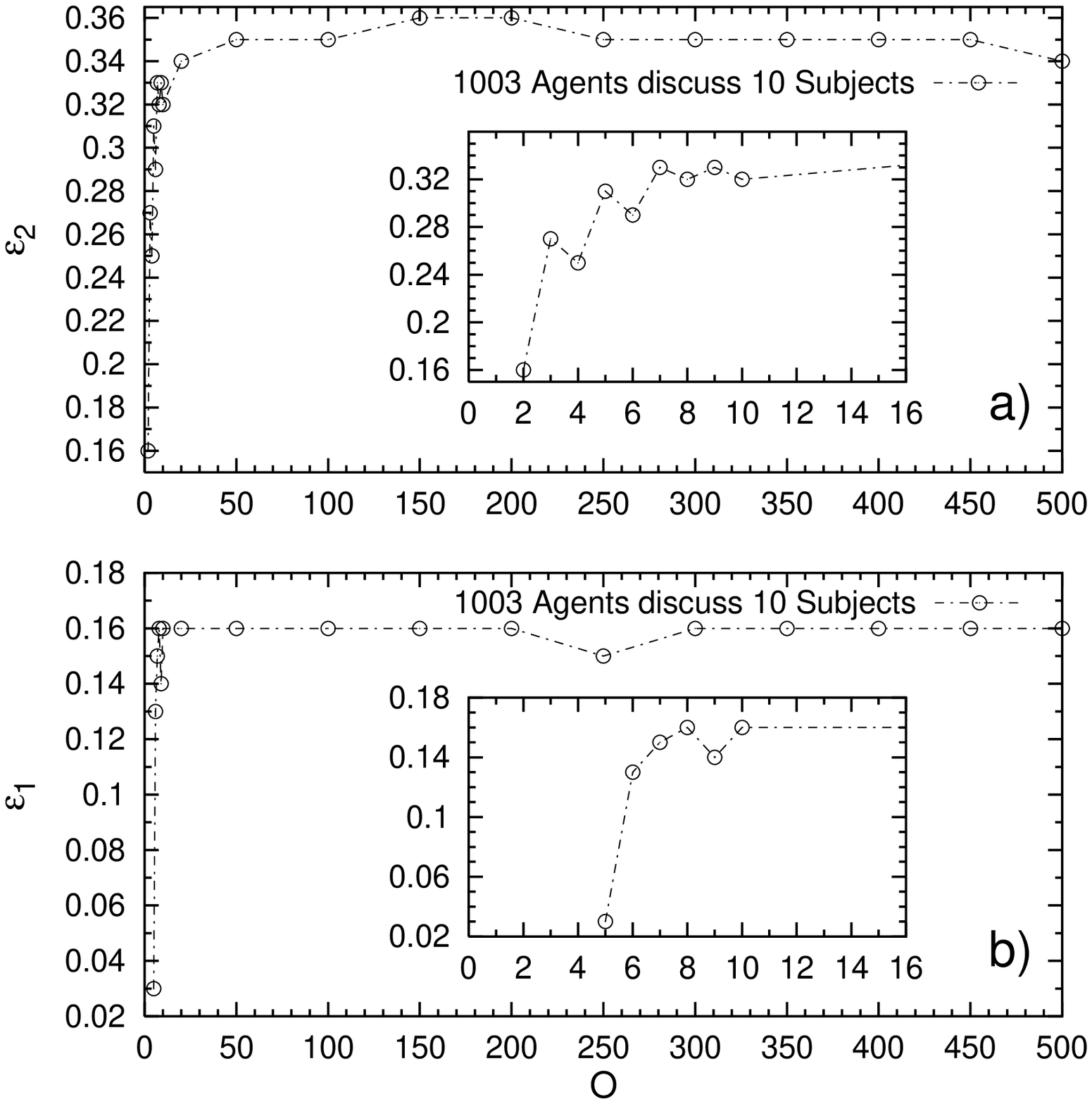, scale=0.7, angle=-90}}
\fcaption{The inset shows the left end more detailed.\\
\hspace*{0.97cm}a) shows that large opinion spectra saturate early
and so only small $O$ have an \\\hspace*{1.22cm} an influence on 
$\varepsilon_2$.\\
\hspace*{0.97cm}b) shows the same for $\varepsilon_1$.}
\label{Ovar}
\end{figure}

\noindent
The opinion spectrum $O$ has only an influence when it is small.
Large spectra saturate very fast. 
$O$ has less influence on $\varepsilon_1$ (Fig.\,\ref{Ovar}\,b{\small )} 
(Inset)) and a little more influence on $\varepsilon_2$.
Increasing $O$ increases both $\varepsilon_1$ and $\varepsilon_2$.\\
Thus diversity of opinion complicates the process of agreement.

\section{No absolute Consensus}
\noindent
Almost always several clusters remain 
of different size (\,= number of agents within one cluster)
in the area $\,>\varepsilon_2$ 
where the number of clusters in the average stays stable small.    

\subsection{Remaining Cluster}
\begin{figure}[ht]
\centerline{\psfig{file=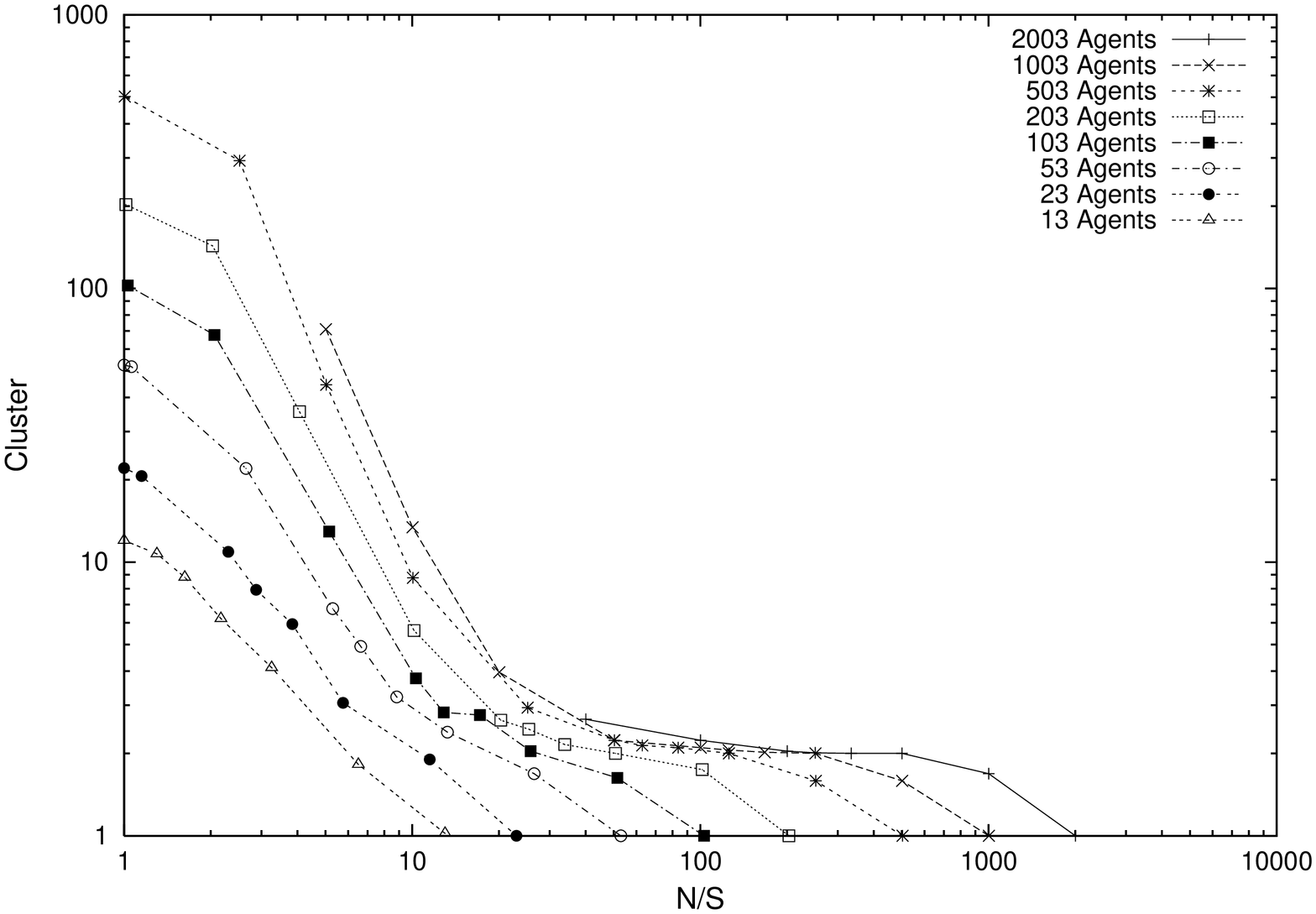, 
scale=0.52, angle=-90}}
\fcaption{Dependence of the number of clusters on the fraction $N$/$S$. 
Over a wide range
about 2 clusters occur. One dominating with a very high
occupation and a 2nd minor, with a low occupation dependent on $S$. 
On the left a region with a rapid growing number of clusters is attached;
many little clusters exist; there is no all-dominant large cluster. 
For $N/S\,<\,1$ the results are the same as for $N/S\,=\,1$.}
\label{Cluster}
\end{figure}

\noindent
The number of remaining clusters depends on the ratio $N/S$ and the 
absolute value of $S$ (Fig.\,\ref{Cluster}).
\begin{itemize}
\item At $S$=1 exists an absolute consensus (remaining cluster = 1).
\item If the ratio $N/S$ is large (small $S$) on average $\simeq$\,2 clusters
remain in the area$\,> \varepsilon_2$. The difference of the clusters is
$\delta\,$=\,1. One cluster dominates in its number of members, the other
depends in its size on $S$. The latter I call ``2nd Cluster''.    
\item When $S$ becomes larger and the ratio $N/S$ smaller, but still
$N/S\,>\,1$, a huge increase
of the number of remaining clusters is seen. At the beginning  there is
still a large central cluster and some few small clusters.
With increasing $S$ an increasing number of differently sized clusters exists.
\item At very large $S$ ($N/S\le$\,1) a great many number
(maximum $N$) of very small clusters exists.
The number of clusters is for $N/S\,<\,1$ the maximum ($N$).
\end{itemize}
If only one subject will be discussed, it gives an absolute consensus.
If several subjects are discussed, no absolute consensus will
be given, but in general a majority opinion.\\
If the ratio of the number of agents to the number of subjects is small, but 
$S\,<\,N$, then a fragmentation into many clusters is seen.

\subsection{Number of Occupants of the 2nd Cluster}
\begin{figure}[ht!]
\centerline{\psfig{file=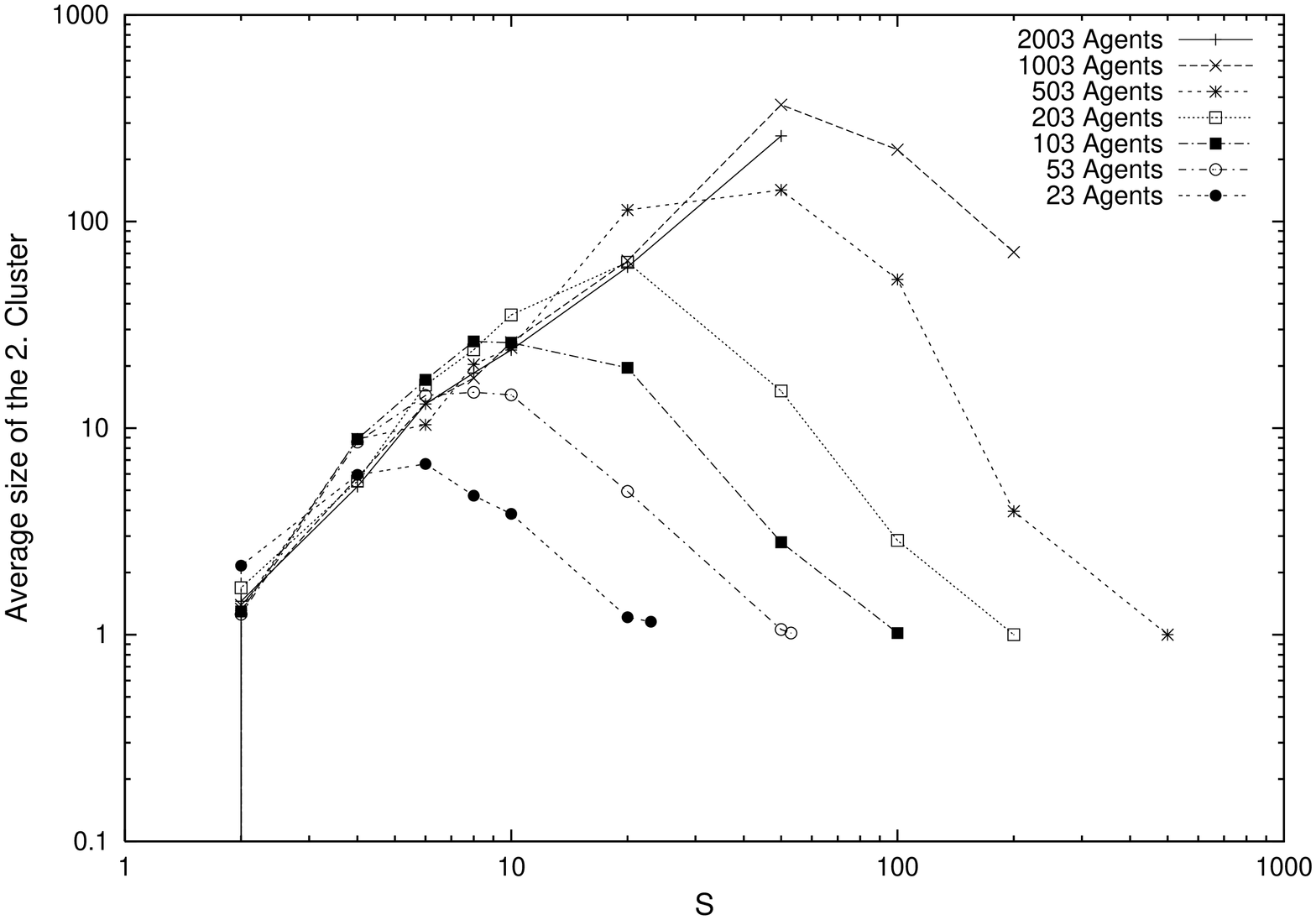, 
scale=0.52, angle=-90}}
\fcaption{The maximum in the number of the size of the 2nd cluster 
is reached in the area where the number of cluster increases with 
increasing $S$. In the extreme areas of many existing
clusters and absolute consensus respectively, the number of 
occupants decreases to 1 resp.\,0.}
\label{Besetzung}
\end{figure}

\begin{figure}[ht!]
\centerline{\psfig{file=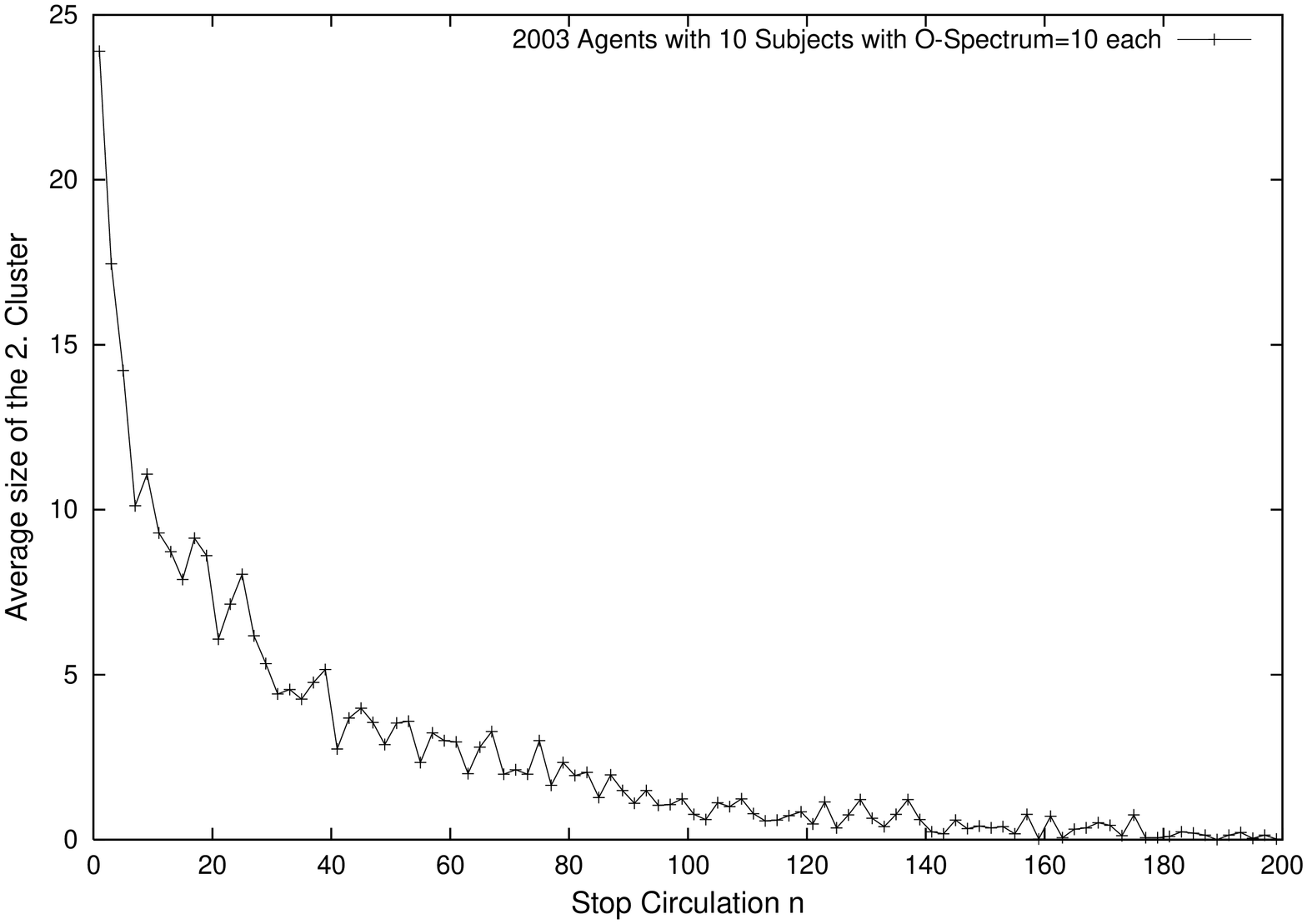, scale=0.52, angle=-90}}
\fcaption{The existence of the 2nd cluster stays highly robust. Even after
$n$\,=\,200 consecutive iterations of communication there still remain 
occupants.}
\label{Stop}
\end{figure}

\noindent
The case of two remaining cluster is prevalent over a wide range of
$N/S$. Therefore I have considered particularly the occupation of the
2nd cluster.\\
The size of the 2nd cluster varies proportional to
$S$ (Fig.\,\ref{Besetzung}) for intermediate $S$.
Even a rise of $n$ to 200 iterations 
does not erase the existence of the 2nd cluster (Fig.\,\ref{Stop}).

\subsection{Explanation}
\noindent
\begin{figure}[ht!]
\centerline{\psfig{file=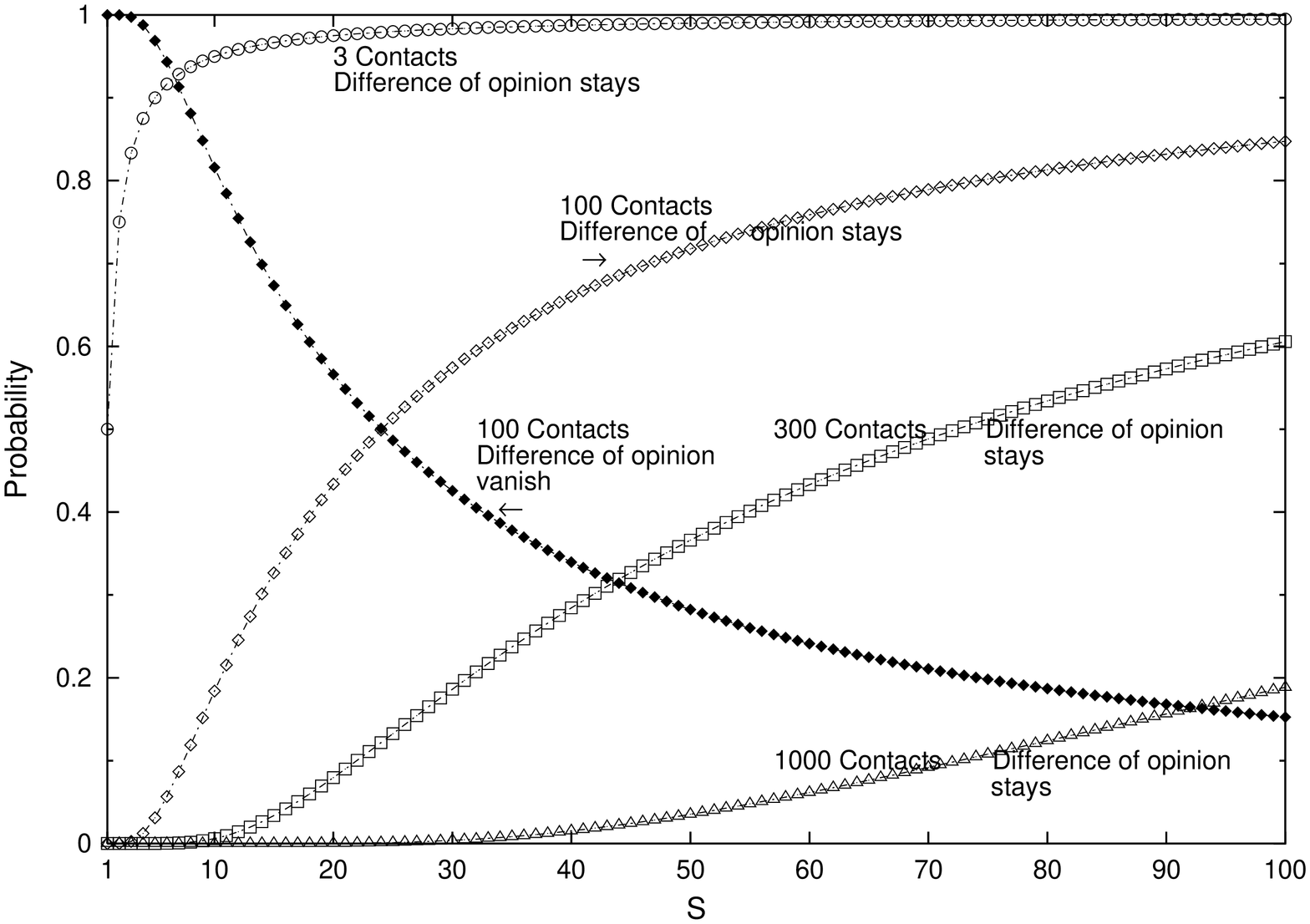, 
scale=0.52, angle=-90}}
\fcaption{The Probability for an existing difference in opinions 
to stay depends on the number of subjects $S$ and on how often
the agent with the opinion at issue will be spoken to. This depends on
the number of its connections $k$ within the network. Many connections 
result in many addresses and therefore in a low possibility of surviving
for a opinion at issue.\\
This graph shows the equation \ref{math3} for different 
$k$\,=\,3, 100, 300, 1000 and once the equation \ref{math4} 
for $k$\,=\,100 (full diamonds).}
\label{Barriere}
\end{figure}
I have looked for an explanation of the effects described above.\\
A variation of this model has been simulated, in which
one subject after the other is simulated. The whole population discussed the 
same subject until the stop criterion for this subject is reached.
In this modified model there are always both, an absolute consensus and also
occupied
{\bf p}laces of the {\bf d}istribution with a {\bf i}nter{\bf m}ediate 
number of agreements (:= PDIM, cp. Fig.\,\ref{Vert}).
Consequently the existence of a 2nd (or several) remaining cluster(s)
arises from the random selection of the subject to discuss.\\
Therewith I explain the remaining clusters and their dependence of $S$
as follows:\\
A dynamical barrier may be develop between two agents $i$ and $j$.\\
For explanation I take the state that $i$ and $j$ differ only about
one opinion unit ($\delta$\,=\,1) in subject $S_c$ and agree in all 
other subjects. 
At a discussion between $i$ and $j$ a subject of agreement is
chosen with a probability $1-1/S$. 
Due to the present congruence of opinions the discussion stops immediately.\\  
With probability $1/S$ the subject $S_c$ will be chosen for
discussion. If a discussion will start then the probability is 
$1/2$ that a takeover of a opinion takes place (Second rule for simulating
the communication).
The agents $i$ and $j$ obtain an agreement in subject $S_c$
with probability $\frac{1}{2S}$.\\
Now it is assumed that a consensus of all agents in the network is given 
with only two agents $i$ and $j$ differing in their opinion on
subject $S_c$ by one opinion unit.
The difference in opinion vanishes with a probability 
$\frac{1}{2S}$, leading to total agreement.
Therefore the difference in opinions persists with a probability
$(1\,-\,\frac{1}{2S})$. Depending on
the number $k$ of connections of agent $i$ within the network
the probability is
\begin{equation}
 {\frac{k}{m}\choose e}\,\left(1-\frac{1}{2S}\right)^e\,
\left(\frac{1}{2S}\right)^{\frac{k}{m}-e}
\,,  \label{math2}
\end{equation}
therefore the subject $S_c$ is either not addressed or the partner 
for discussion takes over the opinion of $i$ on the subject $S_c$.
Let $e$ be the number, out of $k/m$ for agent $i$ with $k$ neighbours,
of events where the opinion stays at $S_c$. 
$m$ is the number of agents to talk to, chosen during the assembly 
of the network.
Then the probability for agent $i$ not to change opinion is (2).
Shall the opinion in $S_c$
be preserved then $e$ has to be as large as all $k/m$ requests
for a discussion.
I.e.,
at no call for discussion may the subject $S_c$ be addressed 
or change the opinion. With $e$\,=\,$\frac{k}{m}$ the probabilitiy 
not to change opinion
\begin{equation}
 {\frac{k}{m}\choose \frac{k}{m}}\,\left(1-\frac{1}{2S}\right)^{\frac{k}{m}}\,\left(\frac{1}{2S}\right)^{\frac{k}{m}-\frac{k}{m}}\,\\=\,\left(1-\frac{1}{2S}\right)^{\frac{k}{m}} 
\,,  \label{math3}
\end{equation}
thus is the difference of opinion to stay, and with it
\begin{equation}
 1-\left(1-\frac{1}{2S}\right)^{\frac{k}{m}} 
\,,  \label{math4}
\end{equation}
for the difference to vanish.\\
In this way single nodes could become quasi inactive in relation to
$S_c$. With an existing difference in opinion by $i$ and $j$ at 
subject $S_c$ this difference stays with a probability (3).
If this probability (3) is very small I call it ``dynamic barrier''.\\ 
If the stop criterion intervenes before the dynamic barrier is negotiated
the network owns a node, inactive up to this moment in relation to a subject.\\
Only the same subject can be discussed by two agents. 
So we can imagine, that the same subject together owns a separate network. 
With it, we have $S$ networks. 
Consequently an inactive node in relation to a subject is a inactive
node of a network.\\
Albert et al.(\cite{12}) describe that a not performing node in a 
Barab\'asi-Albert Network leads to clustering.
Thereby a large cluster remains and some small clusters with 
very small number of members (1-16 agents with $m=3$ and $N=10000$).
But this only takes place if no
node with many connections $k$ (center agent) is unperforming.\\
If a center agent (in \cite{12} call Hub) is inactive, then the
network disintegrates in many clusters of different sizes.\\
The probability (3) of a barrier for an agent depending on its $k$
connections can be seen in Figure \ref{Barriere}. A
center agent with many connections has a high probability
for a barrier to vanish.

\subsection{Not occupied PDIM's}
\label{mittlereZ}
\begin{figure}[]
\centerline{\psfig{file=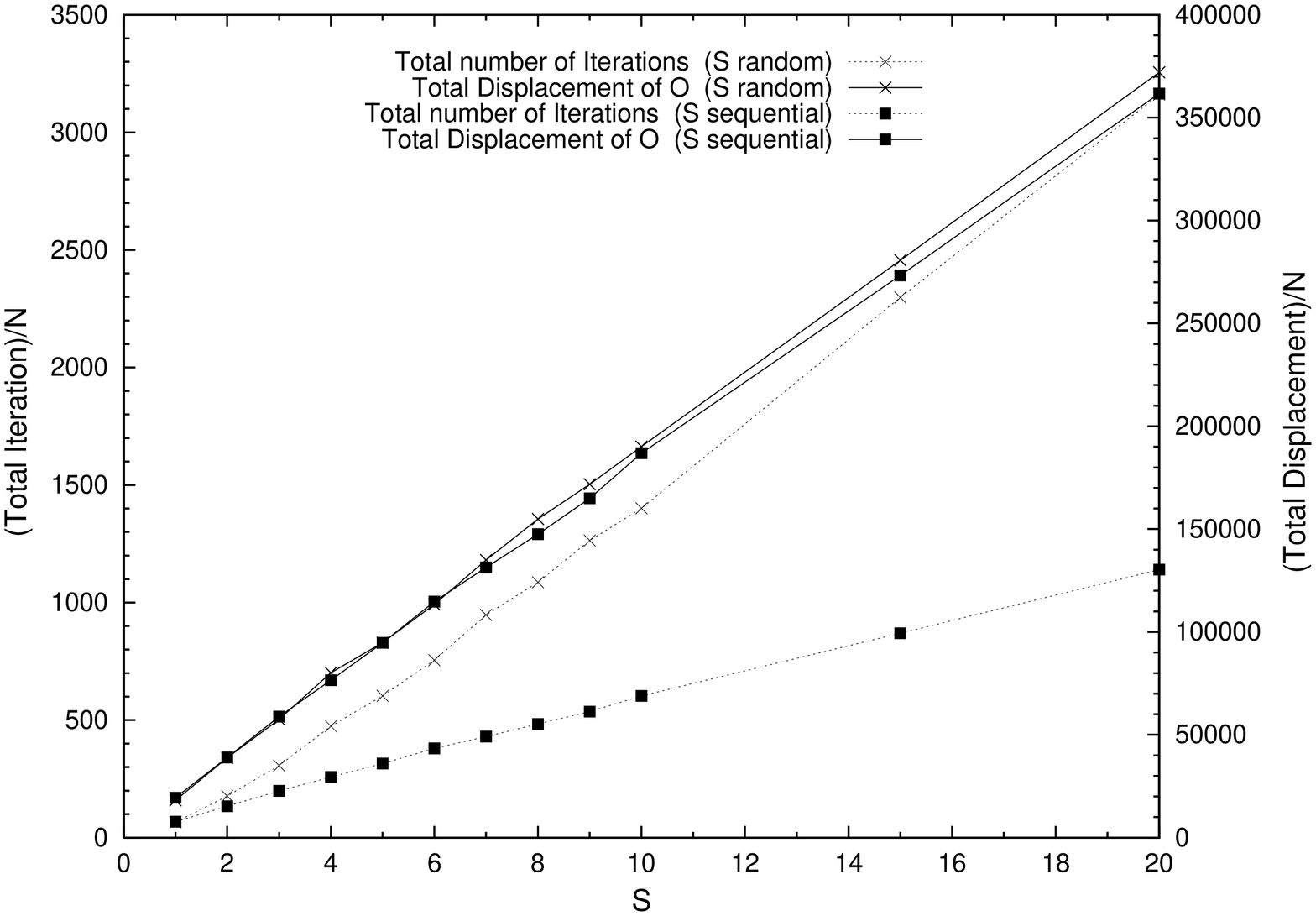, scale=0.52, angle=-90}}
\fcaption{Summed up number of iterations until stop 
and total displacement of opinions (average number of opinion changes)
of a complete simulation for $\varepsilon$=0\,to\,1 
with $\Delta\varepsilon$\,=\,0.01 steps. 
It was regarded a network of $N$\,=\,1003 agents.}
\label{Lauf}
\end{figure}

\noindent
As mentioned before (start of section 4.3), in the distribution 
(Fig.\,\ref{Vert}) the
occupation of the places with an average value of agreed subjects 
depends on the random choice of subjects during the simulation.\\
The non-occupation of these intermediate states 
required that nearly all agents which can discuss with agent $i$ 
at a particular $\varepsilon$, will discussed until consensus is 
reached, before the simulation stops.
If it is not discussed up to a high level of agreement there would 
be given agents which agree in an intermediate number of subjects.\\ 
A discussion up to a near consensus can only happen, if the number of 
iterations during the communication is large enough.
The random (as opposed to sequential at start of section 4.3) choice 
of subject let the number of iterations raise (Fig.\,\ref{Lauf}) in the 
network.
The number of iteration seems to be high enough, that every agent, which 
can and has begun to discuss, will discuss up to the end (discuss till 
a near consensus).

\subsection{General Remarks}\label{All}
\noindent
The agents that occupy the extremes clusters
at first are the center agents. 
Because of their high number of connections $k$, they often get in 
touch with agents which hold many different opinions in their subjects 
and from there are forced into a centrist opinion where they stay.
The stabilised center agents pull the agents they are connected to 
into their state of opinion by their links.\\
Center agents are leaders which are formed and stabilised by public opinion.

\section{Conclusion}
\noindent
The behaviour of this model with its multi-components subjects
on a scale-free network is not analogous to hitherto simulated
one subject-models or models on vectors, lattices or higher
dimensional-geometries.\\
The essential difference to these is the missing of a total consensus
at more than one subject ($S >$\,1). In the present models the number
of remaining cluster is $\propto\,S/N$ (cp. Fig. \ref{Cluster}).

\nonumsection{Acknowledgement}
\noindent
I would like to thank D. Stauffer that he has given me 
the possibility to work on this subject.
And I thanks very much N. Klietsch for stimulating discussions.

\nonumsection{References}

\end{document}